\documentclass[a4paper,11pt]{article}
\usepackage{pos}

\title{UPCs as probes of partonic structure -- exclusive and inclusive processes}
\author*[a,b]{Vadim Guzey}

\affiliation[a]{University of Jyv\"askyl\"a, Department of Physics, P.O. Box 35, FI-40014 University of Jyv\"askyl\"a, Finland}

\affiliation[b]{Helsinki Institute of Physics, P.O. Box 64, FI-00014 University of Helsiniki, Finland}

\emailAdd{vadim.a.guzey@jyu.fi}

\abstract{
Ultraperipheral collisions (UPCs) at the LHC and RHIC provide important new information on the partonic structure of the proton and nuclei and
small-$x$ dynamics in QCD. We review phenomenological applications of the 
collinear factorization at leading and next-to-leading orders of perturbative QCD and the dipole model to coherent and incoherent 
$J/\psi$ photoproduction in Pb-Pb UPCs at the LHC emphasizing the strong leading twist gluon nuclear shadowing, the role of quark-antiquark-gluon dipoles, and a possible onset the gluon saturation in nuclei. We also discuss inclusive and diffractive dijet photoproduction in UPCs, which give complementary constraints
on nuclear parton distributions and the pattern of factorization breaking in 
diffraction.

}

\FullConference{HardProbes2023\\
 26-31 March 2023\\
 Aschaffenburg, Germany\\}

\begin{document}
\maketitle

\section{Ultraperipheral collisions as a proxy for photon-nucleus collider}
\label{sec:intro}

Ultraperipheral collisions (UPCs) constitute an important part of the nuclear physics program at the Large Hadron Collider (LHC) and Relativistic Heavy Ion Collider (RHIC).
In UPCs, colliding ions pass each other at large impact parameters of the order of tens of fermi. In this case, the strong hadron-hadron interaction is suppressed and the reaction is mediated by quasi-real photons in the Weizs\"acker-Williams equivalent photon approximation. 
The flux of these photons is enhanced by $Z^2$ ($Z$ is the ion electric charge) and the maximal photon energy is proportional to the ion Lorentz factor $\gamma_L$. This makes UPCs a proxy for a high-energy photon-nucleus collider reaching at the LHC well into the TeV-range.
Thus, UPCs give an opportunity to address open questions of the partonic structure of protons and nuclei and the strong interaction dynamics in small-$x$ QCD~\cite{Bertulani:2005ru,Baltz:2007kq,Contreras:2015dqa,Klein:2019qfb}.

One of the most frequently studied UPC process is photoproduction of light ($\rho)$ and heavy ($J/\psi$, $\Upsilon$) vector mesons $V$. Since both colliding nuclei can serve as a source of
photons and as a target, the UPC cross section is given by a sum of two terms,
\begin{equation}
\frac{d\sigma^{AA \to VAA^{\prime}}}{dy dp_T}=\left[kN_{\gamma/A}(k) \frac{d\sigma^{\gamma A \to V A^{\prime}}}{dp_T}\right]_{k=k^{+}}+\left[kN_{\gamma/A}(k) \frac{d\sigma^{\gamma A \to V A^{\prime}}}{dp_T}\right]_{k=k^{-}} \,,
\label{eq:upc}
\end{equation}
where $y$ and $p_T$ are the vector meson rapidity and transverse momentum, respectively;
$kN_{\gamma/A}(k)$ is the photon flux known from QED with
an additional input to suppress the strong interaction at small impact parameters.
Equation~(\ref{eq:upc}) leads to a two-fold ambiguity in the photon energy $k$ as a function of $y$,
$k^{\pm}=(M_V/2) e^{\pm y}$, where $M_V$ is the vector meson mass.
The underlying photon-nucleus cross section  $d\sigma^{\gamma A \to V A^{\prime}}/dp_T$ can be either coherent ($A^{\prime}=A$, target intact) or incoherent ($A^{\prime} \neq A$, target breaks up). The two cases contain complementary information on small-$x$ dynamics and are experimentally distinguished by measuring the $p_T$ distribution and comparing it to predictions of the STARlight Monte Carlo~\cite{Klein:2016yzr}.

Both coherent and incoherent scattering can be accompanied by mutual electromagnetic excitation of the colliding ions with their subsequent de-excitation and emission of forward neutrons~\cite{Pshenichnov:2001qd,Baltz:2002pp}.
Measurements of UPCs in conjunction with the detection of forward neutrons in any two channels (0n0n, 0nXn, XnXn) allow one to circumvent the photon-energy ambiguity and thus to access very low values of $x$~\cite{Guzey:2013jaa,Kryshen:2023bxy}.
In the case of coherent $J/\psi$ photoproduction in Pb-Pb UPCs at 5.02 TeV, this was recently demonstrated by
CMS~\cite{CMS:2023snh} and ALICE~\cite{ALICE:2023jgu} collaborations at the LHC.

\section{Exclusive $J/\psi$ photoproduction in collinear factorization of pQCD}
\label{sec:collinear}

Studies of exclusive $J/\psi$ photoproduction in UPCs are motivated by the result that in the leading double logarithmic approximation of perturbative QCD (pQCD) and the static limit for the charmonium wave function, the cross section of this process is proportional to the small-$x$ gluon density of the target squared~\cite{Ryskin:1992ui},
\begin{equation}
 \frac{d\sigma^{\gamma p \to J/\psi p}(t=0)}{dt} = \frac{12 \pi^3}{\alpha_{\rm e.m.}}  \frac{\Gamma_V M_{V}^3 }{( 4 m_c^2)^4}\left[\alpha_s(Q_{\rm eff}^2) xg(x,Q_{\rm eff}^2)\right]^2 C(Q^2=0) \,,
 \label{eq:jpsi}
\end{equation}
where $\Gamma_V$ is the $J/\psi \to l^{+} l^{-}$ leptonic decay width, $x=M_{J/\psi}^2/W_{\gamma p}^2$ and  $Q_{\rm eff}={\cal O}(m_c)$ with  $W_{\gamma p}$ being the 
photon-nucleon center-of-mass energy and
$m_c$ the charm quark mass. The factor of $C(Q^2=0)$ accounts for effects beyond the non-relativistic approximation for the $J/\psi$ vertex. 

Applying Eq.~(\ref{eq:jpsi})  to nuclear targets, one can express the $p_T$-integrated cross section of coherent $J/\psi$ photoproduction in the following form,
\begin{equation}
\sigma^{\gamma A \to J/\psi A}(W_{\gamma p})=\frac{d\sigma^{\gamma p \to J/\psi p}(t=0)}{dt} \left[\frac{xg_A(x,Q_{\rm eff}^2)}{A xg_p(x,Q_{\rm eff}^2)}\right]^2 \int_{|t_{\rm min}|}^{\infty} dt |F_A(t)|^2 \,,
\label{eq:jpsi2}
\end{equation}
where $xg_A/(A xg_p)$ is ratio of the gluon density in the nucleus to that in the 
free proton, and $F_A(t)$ is the nucleus form factor. 
To quantify the magnitude of nuclear modifications, it is convenient to introduce the nuclear suppression factor $S_{Pb}(x)$~\cite{Guzey:2013xba,Guzey:2013qza},
\begin{equation}
S_{Pb}(x)=\left[\frac{\sigma^{\gamma A \to J/\psi A}(W_{\gamma p})}{\sigma_{\rm IA}^{\gamma A \to J/\psi A}(W_{\gamma p})} \right]^{1/2} =\frac{xg_A(x,Q_{\rm eff}^2)}{A xg_p(x,Q_{\rm eff}^2)} =R_g(x,Q_{\rm eff}^2)\,, 
\label{eq:S}
\end{equation}
where $\sigma_{\rm IA}^{\gamma A \to J/\psi A}$ is the impulse approximation (IA) for the cross section in Eq.~(\ref{eq:jpsi2}), where one neglects nuclear modifications of $xg_A$.
The values of $S_{Pb}(x)$ extracted from ALICE and CMS measurements 
give direct evidence of the large gluon nuclear shadowing~\cite{Guzey:2013xba,Guzey:2013qza},
\begin{equation}
  R_g(x=6 \times 10^{-4}-10^{-3}, Q_{\rm eff}^2=3\ {\rm GeV}^2) \approx 0.6 \,. 
\end{equation}
It confirmed predictions of the leading twist approximation (LTA) to nuclear shadowing~\cite{Frankfurt:2011cs} and the EPS09 nuclear parton distribution functions (nPDFs)~\cite{Eskola:2009uj}.

One can go beyond the approximation of Eq.~(\ref{eq:jpsi}) and use the formalism of collinear factorization for hard exclusive
processes at next-to-leading order (NLO) accuracy. In this framework, the amplitude of exclusive $J/\psi$ photoproduction 
is expressed in terms of generalized parton distributions (GPDs) and NLO gluon and quark coefficient functions $T_g$ and $T_q$~\cite{Ivanov:2004vd}. The nuclear photoproduction amplitude is given by the following convolution~\cite{Eskola:2022vpi,Eskola:2022vaf}, 
\begin{equation}
{\cal M}^{\gamma A \to J/\psi A}(t) \propto \sqrt{\langle O_1 \rangle_{J/\psi} } \int^{1}_{-1}dx \left[T_g(x,\xi) F_A^g(x,\xi,t)+T_g(x,\xi) F_A^{q,S}(x,\xi,t)\right] \,,   
\end{equation}
where $\langle O_1 \rangle_{J/\psi}$ is the non-relativistic QCD (NRQCD) matrix element related to $\Gamma_V$, and $F_A^g$ and $F_A^{q,S}$ are the nuclear gluon and quark singlet GPDs. While only the gluons contribute at leading order (LO), there is both gluon and quark contributions at NLO.

GPDs interpolate between usual PDFs, distribution amplitudes and elastic form factors and, as a result, depend on the two momentum fractions
$x$ and $\xi \approx M_{J/\psi}^2/(2 W_{\gamma p}^2)$, the momentum transfer $t$ and the factorization scale $\mu_F$ (implicit). However, at small $\xi$, GPDs can be expressed in terms of usual PDFs because the  $\mu_F$ evolution washes out information on the possible $\xi$ dependence of GPDs at the input scale~\cite{Shuvaev:1999ce,Dutrieux:2023qnz}.
Therefore, with a good accuracy, one can use the following model for the gluon nuclear GPD (the quark case is similar),
\begin{equation}
F_A^g(x,\xi,t,\mu_F)=xg_A(x,\mu_F) F_A(t) \,.    
\end{equation}

\begin{figure}[t]
\centerline{%
\includegraphics[width=12.cm]{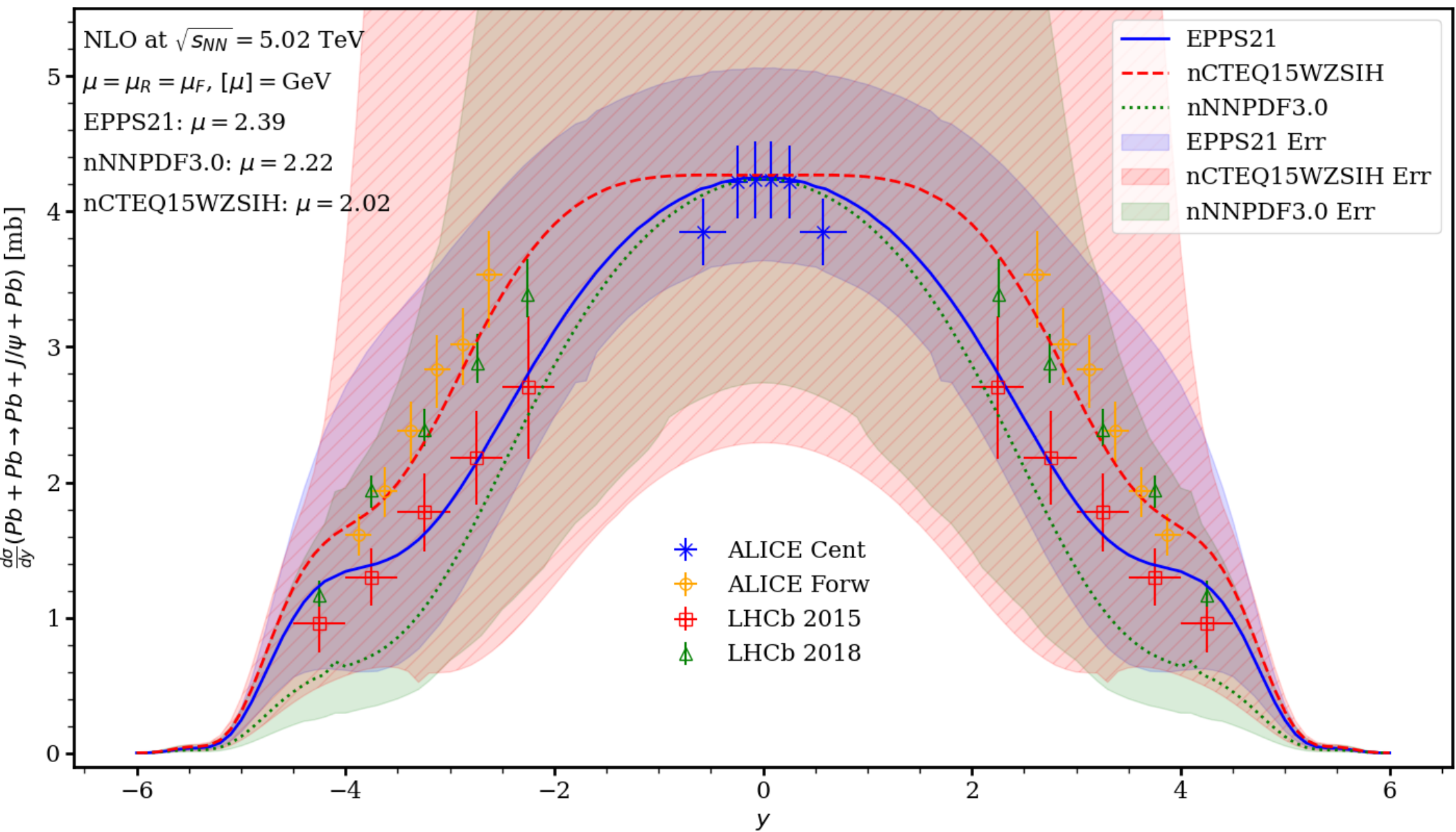}}
\caption{NLO pQCD predictions for 
$d\sigma^{AA\to AA J/\psi}/dy$ as a function of the $J/\psi$ rapidity $y$ in Pb-Pb UPCs at 5.02 TeV for EPPS21, nNNPDF3.0, and nCTEQ15WZSIH nPDFs. For comparison, the Run 2 LHC data are also shown.  From~\cite{Eskola:2022vaf}.}
%vgThe curves correspond to the central nPDFs at the appriate
%optimal scales and the shaded bands show the propagated nPDF uncertainties. For comparison, the Run 2 LHC data are also shown. }
\label{fig:EPPS21_run2_error}
\end{figure}

Figure~\ref{fig:EPPS21_run2_error} shows NLO pQCD predictions for the rapidity dependence of the cross section of coherent $J/\psi$ photoproduction in Pb-Pb UPCs at 5.02 TeV, 
which are compared to the Run 2 LHC data. 
In the calculation~\cite{Eskola:2022vaf}, three state-of-the art nPDFs (EPPS21, nCTEQ15WZHIS, and nNNPDF3.0) have been used. The shaded bands showing the propagation of
the nPDF uncertainties are much larger than the experimental errors, which implies that these data have the potential to improve the determination of nPDFs.

It is well known that NLO pQCD calculations for this process are characterized by a very strong $\mu_F$ dependence, which originates from the double logarithmic terms in the NLO coefficient functions enhanced by $\log(1/\xi)$ at small $\xi$. One can nevertheless find an optimal scale providing a reasonable description of both Run 1 and 2 UPC data (these values are shown in the figure).

A surprising result of the NLO pQCD analysis~\cite{Eskola:2022vpi,Eskola:2022vaf} is the dominance of the quark contribution at central rapidities $y$ because of the strong cancellations between the LO and NLO gluon contributions. It challenges the 
perturbative stability of the NLO calculations and the interpretation of the UPC data in terms of the gluon nuclear shadowing. This can be partially mitigated by considering the ratio of UPC cross sections for two different targets, 
for instance, for oxygen and lead, where the $\mu_F$ dependence is dramatically reduced and becomes comparable to nPDF uncertainties~\cite{Eskola:2022vaf}.

At the same time, NLO pQCD predictions for $\Upsilon$ photoproduction are under better theoretical control than those for $J/\psi$: 
NLO corrections are moderate, GPD modeling benefits from the longer evolution in 
$\mu_F$, and relativistic corrections are expected to be small. 
This was demonstrated in the case of $\Upsilon$ photoproduction in Pb-Pb UPCs at the LHC~\cite{Eskola:2023oos}, which was shown to be dominated by the gluon contribution and which hence can be used to study the $\mu_F$ dependence of the gluon nuclear shadowing.

It was argued that perturbative stability of the NLO predictions for heavy vector meson photoproduction can be improved by a special choice of the factorization scale  $\mu_F=m_c$ and the so-called $Q_0$-subtraction eliminating certain double counting between NLO and LO terms~\cite{Jones:2015nna,Jones:2016ldq}. In the case of the proton target, it was shown to restore the gluon dominance leading to new constraints on the gluon distribution at small $x$~\cite{Flett:2019pux,Flett:2020duk}.

\section{Exclusive $J/\psi$ photoproduction in dipole picture}
\label{sec:dipole}

 In the dipole picture of high-energy scattering, the photon is viewed as a superposition of long-lived quark-antiquark, quark-antiquark-gluon, etc.~dipoles, which elastically scatter on the target nucleons. This leads to
a high-energy factorization for the nuclear scattering amplitude, which is expressed in terms of the overlap of the photon and $J/\psi$ wave functions and the eikonal form for  the dipole cross section $\sigma(r_T)$,
\begin{equation}
 {\cal M}^{\gamma A \to J/\psi A}=2 \int d^2 {\bf r}_T \int \frac{dz}{4 \pi} \int d^2 {\bf b}_T [\Psi^{\ast}_{J/\psi} \Psi_{\gamma}] \left(1-e^{-\frac{1}{2} \sigma(r_T) T_A(b_T)}\right) \,,
 \label{eq:dipole}
\end{equation}
where $r_T$ is the dipole transverse size, $b_T$ is the impact parameter, $z$ is the light-cone momentum fraction shared by the charm quark and antiquark, and 
$T_A(b_T)$ is the nuclear optical density. 
The application of this implementation of the dipole model to coherent $J/\psi$ photoproduction in Pb-Pb UPCs at 2.76 TeV overestimates the data at $y \approx 0$~\cite{Lappi:2013am,Luszczak:2019vdc} 
because nuclear shadowing resulting from scattering of small dipoles with $r_T \sim 0.3$ fm on target nucleons is weak. It is an example of the general observation that successive scattering of small dipoles on nucleons of a nuclear target corresponds to a higher-twist effect~\cite{Frankfurt:2002kd}.

To cure this, one needs to include higher Fock states, in particular, the quark-antiquark-gluon dipoles. The improved, more complete dipole model predictions agree well 
with the LHC data on coherent $J/\psi$ photoproduction in Pb-Pb UPCs~\cite{Luszczak:2021jtr,Kopeliovich:2020has}.
Note that the inclusion higher Fock states allows one to model inelastic (leading-twist) contribution to nuclear shadowing as required by the Glauber-Gribov theory of nuclear shadowing~\cite{Frankfurt:2000tya}.

Instead of building up the nuclear scattering amplitude as the Glauber series, see Eq.~(\ref{eq:dipole}), one can implement nuclear geometry in the initial condition of 
the Balitsky-Kovchegov (BK) equation. This can be interpreted as an onset of the 
gluon saturation in nuclei, but not necessarily in the nucleons. The resulting approach provides a good description of the LHC data on coherent $J/\psi$ photoproduction in Pb-Pb UPCs~\cite{Bendova:2020hbb}.

To summarize, none of the available theoretical approaches describe the $y$-dependence of
coherent $J/\psi$ photoproduction in Pb-Pb UPCs at 5.02 TeV~\cite{ALICE:2021gpt,LHCb:2021bfl}.
The suppression at $y \approx 0$ can be interpreted as the strong leading twist gluon nuclear shadowing in the collinear framework or as an indication of the importance of 
higher Fock states and the gluon saturation in the dipole picture.
Large values of $y$ correspond to small shadowing, where predictions of different models generally converge, but there they are at the border of their applicability.
In this respect, recent measurements of UPCs with neutron emission~\cite{CMS:2023snh,ALICE:2023jgu} provide important complementary information on 
nuclear shadowing in a wide range of $x$, $10^{-5} < x < 0.03$.

Turning to incoherent $J/\psi$ photoproduction in UPCs, one notices that it requires a new sub-nucleon scale, which can be realized in the form of ``hot gluon spots'' and geometric fluctuations of the proton. Implementation of these ideas in the dipole picture 
provides a fair description of the ALICE data~\cite{Mantysaari:2017dwh,Cepila:2017nef}.
The competing explanation is based on the leading twist approximation (LTA) to nuclear shadowing~\cite{Guzey:2018tlk}. 
Note that the two approaches provide compatible predictions for the $t$-dependence
of the $J/\psi$ photo-nuclear cross section, which are characterized by a shift toward smaller $|t|$ in the coherent case~\cite{Guzey:2016qwo,ALICE:2021tyx} and flattening for large $|t|$ in the incoherent case.

\section{Inclusive and diffractive dijet photoproduction in Pb-Pb UPCs}

Complementary information on the partonic structure of nuclei and small-$x$ QCD dynamics can be obtained in dijet photoproduction in UPCs. In the framework of collinear factorization and NLO pQCD, the cross section of inclusive dijet photoproduction in Pb-Pb UPCs reads~\cite{Guzey:2018dlm}
\begin{equation}
 d\sigma^{AA \to A+{\rm 2jets}+X}=\sum_{a,b}\int dy \int dx_{\gamma} \int dx_A f_{\gamma/A}(y)f_{a/\gamma}(x_{\gamma},\mu^2)f_{b/A}(x_A,\mu^2) {\hat \sigma}_{ab \to {\rm jets}} \,,
 \label{eq:jets}
\end{equation}
where $f_{\gamma/A}(y)$ is the photon flux, $f_{a/\gamma}(x_{\gamma})$ are photon PDFs in the resolved-photon case, $f_{b/A}(x_A)$ are nuclear PDFs, and  ${\hat \sigma}_{ab \to {\rm jets}}$ is the cross section to produce jets in hard scattering of partons $a$ and $b$ (in the direct photon case, $a$ stands for the photon). The distributions in Eq.~(\ref{eq:jets}) depend on the corresponding momentum fraction and the factorization scale $\mu$.  

\begin{figure}[t]
\centerline{%
\includegraphics[width=10.cm]{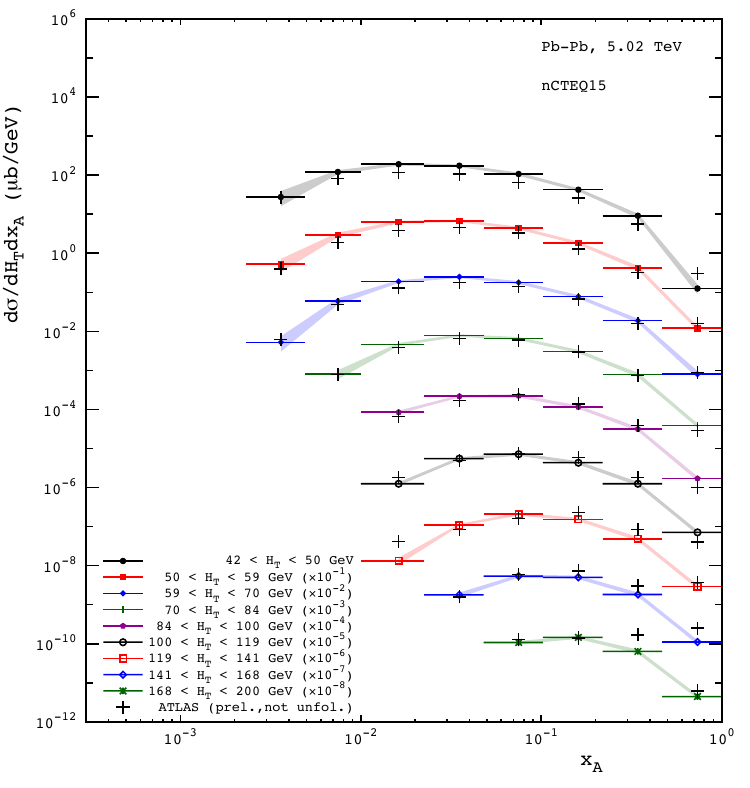}}
\caption{NLO pQCD predictions for the cross section of inclusive dijet photoproduction in Pb-Pb UPC as a function of $x_A$ in different bins of the dijet transverse momentum $H_T$ vs.~the ATLAS data.  From~\cite{Guzey:2018dlm}.}
\label{fig:jets}
\end{figure}

Figure~\ref{fig:jets} shows predictions of Eq.~(\ref{eq:jets}) with the implementation and input specified in~\cite{Guzey:2018dlm} and compares them to the preliminary ATLAS data at 5.02 TeV~\cite{ATLAS:2017kwa}. One can see from the figure that NLO pQCD describes well the shape and normalization of the data.
This cross section is sensitive to nuclear modifications of nPDFs at the level of 
$10-20$\% and can be used to reduce their uncertainties by approximately a factor 
of 2~\cite{Guzey:2019kik}. 
Note that this process can also be used to look for nonlinear effects within the 
color glass condenstate (CGC) framework~\cite{Kotko:2017oxg}.

By requiring that the target nucleus stays intact, one study diffractive dijet photoproduction in UPCs. In NLO pQCD, the $k^{+}$ contribution [see Eq.~(\ref{eq:upc})] to the cross section of diffractive dijet photoproduction in Pb-Pb UPCs has the following form~\cite{Guzey:2016tek},
\begin{eqnarray}
 d\sigma^{AA \to A+{\rm 2jets}+X+A(+)}&=&\sum_{a,b}\int dy \int dx_{\gamma} \int dx_{P} \int dz_P \nonumber\\
 &\times& 
 f_{\gamma/A}(y)f_{a/\gamma}(x_{\gamma},\mu^2)f_{b/A}^{D(3)}(x_P,z_P,\mu^2) {\hat \sigma}_{ab \to {\rm jets}} \,,
 \label{eq:jets2}
\end{eqnarray}
where $x_P$ and $z_P$ refer to the Pomeron-in-nucleus and parton-in-Pomeron momentum fractions.
Such measurements will probe the novel nuclear diffractive PDFs $f_{b/A}^{D(3)}$ at small $x$~\cite{Frankfurt:2011cs} and may also shed some light on the mechanism of factorization breaking in 
diffractive scattering~\cite{Guzey:2016tek}.

\section{Summary}

There is continuing interest in using UPCs at the LHC and RHIC to obtain new constraints on the partonic structure of the proton and nuclei and
the QCD dynamics at small $x$. 
The available data challenge both collinear factorization and dipole model frameworks. 
Large nuclear suppression of coherent $J/\psi$ photoproduction in Pb-Pb
UPCs at the LHC  can be explained by the strong gluon/quark shadowing at small $x$, 
the contribution of quark-antiquark-gluon dipoles, or an onset of saturation.
It will be important to test these prediction in $\Upsilon$ photoproduction in Pb-Pb UPCs.

While the framework of collinear factorization and NLO pQCD provides a fair 
description of coherent $J/\psi$ photoproduction in Pb-Pb UPCs at the LHC within large factorization scale and nPDF uncertainties, there remain outstanding 
challenges related to NRQCD corrections to the $J/\psi$ vertex and small-$x$ resummation of the NLO coefficient functions. One way to circumvent these issues is to consider the ratio of UPC cross sections for different targets, where 
the theoretical uncertainties are significantly reduced.
Recent progress in calculations of exclusive vector meson prodiction at NLO in the dipole picture should help to clarify interpretation of the UPC data.

\acknowledgments

This research was funded by the Academy of Finland project
330448, the Center of Excellence in Quark Matter of the Academy of Finland
(projects 346325 and 346326), and the European Research Council project
ERC-2018-ADG-835105 YoctoLHC.

\bibliographystyle{JHEP}
\bibliography{my_bib}

%\begin{thebibliography}{99}

%\cite{Bertulani:2005ru}
%\bibitem{Bertulani:2005ru}
%C.~A.~Bertulani, S.~R.~Klein and J.~Nystrand,
%\emph{Physics of ultra-peripheral nuclear collisions}, 
%Ann. Rev. Nucl. Part. Sci. \textbf{55} (2005), 271-310
%doi:10.1146/annurev.nucl.55.090704.151526
%[arXiv:nucl-ex/0502005 [nucl-ex]].
%452 citations counted in INSPIRE as of 11 Jul 2023

%\end{thebibliography}

\end{document}